\documentclass[preprint,aip,apl,amsmath,amssymb]{revtex4-1}
\usepackage{amsmath}    
\usepackage{graphicx}   
\usepackage{color}
\usepackage[10pt]{moresize}

\usepackage{amsfonts}
\usepackage{amssymb}
\usepackage{amscd}
\usepackage{enumerate}
\usepackage{epsfig}
\usepackage{subfigure}
\usepackage{graphicx}
\usepackage{bm}

\begin{document}

\title{Practical and fast quantum random number generation based on photon arrival time relative to external reference}
\author{You-Qi Nie}
\affiliation{Hefei National Laboratory for Physical Sciences at the Microscale and Department of Modern
Physics, University of Science and Technology of China, Hefei, Anhui 230026, China}
\affiliation{Synergetic Innovation Center of Quantum Information and Quantum Physics,
University of Science and Technology of China, Hefei, Anhui 230026, China}
\author{Hong-Fei Zhang}
\affiliation{Hefei National Laboratory for Physical Sciences at the Microscale and Department of Modern
Physics, University of Science and Technology of China, Hefei, Anhui 230026, China}
\author{Zhen Zhang}
\affiliation{Center for Quantum Information, Institute for Interdisciplinary Information Sciences,
Tsinghua University, Beijing 100084, China}
\author{Jian Wang}
\affiliation{Hefei National Laboratory for Physical Sciences at the Microscale and Department of Modern
Physics, University of Science and Technology of China, Hefei, Anhui 230026, China}
\author{Xiongfeng Ma}
\affiliation{Center for Quantum Information, Institute for Interdisciplinary Information Sciences,
Tsinghua University, Beijing 100084, China}
\affiliation{Synergetic Innovation Center of Quantum Information and Quantum Physics,
University of Science and Technology of China, Hefei, Anhui 230026, China}
\author{Jun Zhang}
\email{zhangjun@ustc.edu.cn}
\affiliation{Hefei National Laboratory for Physical Sciences at the Microscale and Department of Modern
Physics, University of Science and Technology of China, Hefei, Anhui 230026, China}
\affiliation{Synergetic Innovation Center of Quantum Information and Quantum Physics,
University of Science and Technology of China, Hefei, Anhui 230026, China}
\author{Jian-Wei Pan}
\affiliation{Hefei National Laboratory for Physical Sciences at the Microscale and Department of Modern
Physics, University of Science and Technology of China, Hefei, Anhui 230026, China}
\affiliation{Synergetic Innovation Center of Quantum Information and Quantum Physics,
University of Science and Technology of China, Hefei, Anhui 230026, China}
\date{\today}

\begin{abstract}
We present a practical high-speed quantum random number generator, where the timing of single-photon detection relative to an external time reference is measured as the raw data. The bias of the raw data can be substantially reduced compared with the previous realizations. The raw random bit rate of our generator can reach 109 Mbps. We develop a model for the generator and evaluate the min-entropy of the raw data. Toeplitz matrix hashing is applied for randomness extraction, after which the final random bits are able to pass the standard randomness tests.
\end{abstract}

\maketitle

Random numbers have a wide range of applications in our daily life, such as encryption, Monte Carlo simulation, statistical analysis and lottery.
There are mainly two kinds of random number generators, pseudo-random number generators (PRNGs) and true random number generators (TRNGs).
A PRNG generates bitstreams based on deterministic algorithms, which may need an initial random seed as its input.
Since the entropy cannot increase via deterministic algorithms, the generated numbers are not truly random,
which restricts its application in certain fields, especially those in cryptographic usage.

The randomness from a TRNG, on the other hand, is based on fundamental principles of physics.
The indeterministic nature of quantum mechanics allows us to construct quantum random number generators (QRNGs)
whose output cannot be predicted. With appropriate modeling, one can prove the generated bits to be
information-theoretically random after certain data post-processing called randomness extraction.
Many QRNG schemes have been proposed in the past decade based on a variety of principles.

A natural way to construct QRNGs is based on the random path selection of single photons arriving at a 50/50 beam splitter (BS) \cite{ROT94,SGGZ00,JAWWZ00}.
A single-photon detector is used at each output port of the BS to detect which path single photons pass from.
This kind of QRNG can generate at most one random bit per detection event, e.g., bit `0' for one port and bit `1' for the other.
As a consequence, the speed of this type of QRNGs is limited by the detector count rate.
The final random number generation rate is less than one bit per detection event.
Moreover, the generated raw random bits are usually biased due to device imperfections such as unbalanced split ratio and detection efficiency mismatch.

\begin{figure}[t]
\centering
\includegraphics[width=8 cm]{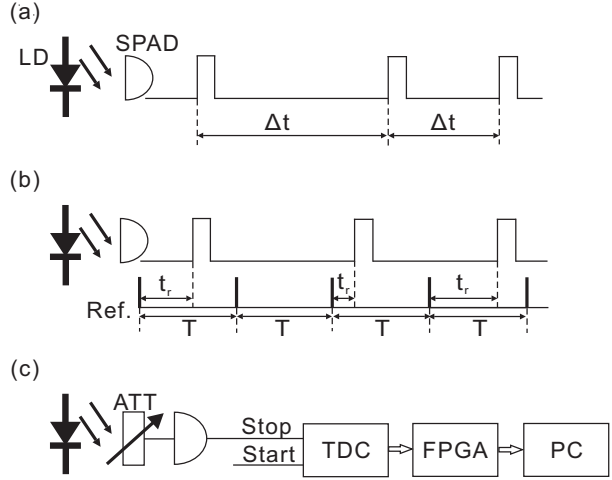}\\
\caption{(a) The previous QRNG schemes by measuring the time intervals between successive photons. (b) Our QRNG scheme using an external time reference. (c) Experimental setup of our scheme.} \label{fig1}
\end{figure}

Recently, the measurement of photon arrival time has been explored as a method of generating high-speed random numbers \cite{USP06,USP07,WJAK09,WK10,WLBRRB11,MXW05,LWWMZ13}.
In this type of QRNGs, as shown in Fig.~\ref{fig1}(a), photons emitted from a continuous-wave (CW) laser diode (LD) are measured by a single-photon detector
and the time intervals ($\Delta t$) between successive detection events are recorded as the raw data.
One detection event can in principle generate $n$ random bits, where $n$ depends on the time resolution of the measurement.
Compared to the aforementioned BS based QRNG scheme whose generation rate is limited by the detector count rate,
the speed of this photon arrival time based QRNG scheme can reach $n$ times higher.
However, a drawback of this scheme is that $\Delta t$ follows an exponential-family distribution and hence the raw random bits are highly biased. In fact, the min-entropy of raw data, which is widely used to quantify the randomness, is relatively small in practice \cite{WJAK09,WLBRRB11,LWWMZ13}.
Therefore, the obtained random bit generation rate would be substantially reduced due to the large bias in the raw data.
To remove such bias, various software-based post-processing methods, such as hash function \cite{WJAK09}, resilient function \cite{WLBRRB11},
and discretized encoding \cite{LWWMZ13}, have been used. Hardware-based schemes are also employed to remove the bias due to the distribution of $\Delta t$. For example, by carefully shaping the laser pulses, $\Delta t$ can approximately follow a uniform distribution \cite{WK10}. Such a hardware fix would increase the complexity of the system. Nevertheless, all these methods are essentially partial randomness extraction, which can be done in post-processing. 

To solve this issue, we present a QRNG scheme based on photon arrival time measurement that takes advantage of an external time reference.
As shown in Fig.~\ref{fig1}(b), a single-photon detector is used to detect photons emitted from a highly attenuated CW laser. The time
difference ($t_r$) between photon detection and an external time reference is measured as the raw data.
Later we will show that the distribution of $t_r$ is approximately uniform in practical implementation and hence the min-entropy of the raw data is close to 1.

The randomness of the photon arrival time stems from the photon number distribution of the CW laser in a given time period, $(t,t+T)$.
The photon number ($k$) follows a Poisson distribution, with the mean photon number of $\lambda T$, where $\lambda$ characterizes the laser intensity,
\begin{equation} \label{pn1}
P(k)=\frac{e^{-\lambda T}(\lambda T)^k}{k!}.
\end{equation}
As shown in Fig.~\ref{fig1}(b), we set the time period of the external reference to be $T$ and divide each period
$(t,t+T)$ into $N_b$ small time bins $\{\tau_1,\tau_2,...,\tau_{N_b}\}$, where $\tau_i=(t+\frac{i-1}{N_b}T,t+\frac{i}{N_b}T)$.
Given a photon detected in a time period $(t,t+T)$, it can be shown that the photon appears in each small time bin $\tau_i$
with the same probability of $1/N_b$. Device imperfections, such as multi-photon emission, detector dead time and detector dark counts,
would degrade the randomness and hence lower the min-entropy of the raw data.
All these effects will be taken into consideration in the randomness extraction.

The experimental setup of our QRNG is shown in Fig.~\ref{fig1}(c). The CW laser is attenuated by a variable optical attenuator (ATT, DA-100).
The laser intensity is chosen so that less than one photon detection happens in each time period $(t,t+T)$ on average.
The photons are detected by a silicon single-photon avalanche diode (SPAD, id100-SMF20-STD),
which has a dead time of 45 ns and a maximum count rate of 13.9 Mcps for continuous light illumination.

To measure the timing of photon detections, we design high-performance timing measurement electronics
with 160 ps time resolution using a time-to-digital converter (TDC) chip (TDC-GPX).
An external time reference is used as the ``start'' of the TDC and the detection signal of SPAD is used as the ``stop''.
The TDC output is fed into a high-speed field-programmable gate array (FPGA, Altera Stratix) and then
the raw random bits are read into a personal computer (PC) via USB2.0.

\begin{figure} [t]
\centering
\includegraphics[width=8 cm]{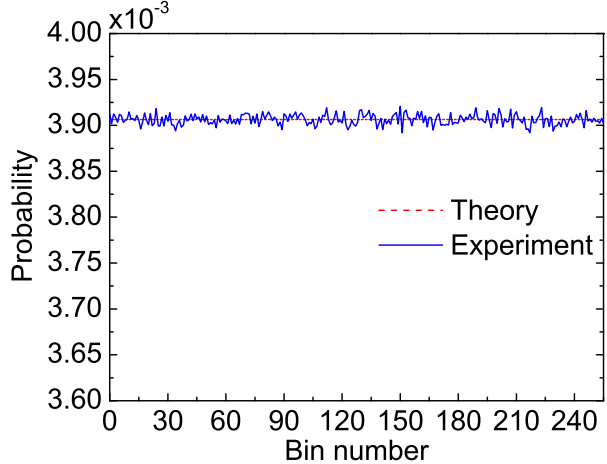}
\caption{Experimental (solid line) versus theoretical (dashed line) probability distribution in 256 bins for 800 Mb raw data.} \label{fig2}
\end{figure}

The raw data from our QRNG follows a uniform distribution very closely, as shown in Fig.~\ref{fig2}.
For an ideal uniform distribution, the probability for a photon detection falling into each of $N_b$ bins ($N_b=T/t_b$) is $1/N_b$.
In our experiment, the number of bins is $N_b=256$ and the time reference period is $T=40.96$ ns.
In the test, we take 800 Mb raw data and calculate the probability in each time bin.
The result is compared with the theoretical value of $1/256$, as shown in Fig.~\ref{fig2}.
One can see that the raw data has a very good quality of randomness. This observation will be confirmed by later post-processing
where the min-entropy of the raw data is evaluated to be close to 1.

To quantitatively evaluate the randomness of the raw data, we need to model the system carefully and figure out the facts that would introduce bias.
There are a few major device imperfections to be examined.
\begin{enumerate}
\item
Detector efficiency, $\eta$, is not unity. We model the efficiency by a BS followed by a perfect detector.
Equivalently, the detector efficiency can be viewed as part of the source intensity attenuation.
Thus, the average photon number $\lambda T$ in Eq.~\eqref{pn1} should be replaced by $\lambda T \eta$.

\item
Detector dark counts may introduce noise. In the experiment, the dark count rate is about 15 cps.
Comparing to the detection count rate of 13.9 Mcps, the effect of dark count is negligible.

\item
The SPAD has a dead time of $\tau_d$. Dead time is a period of time that a detector is inactive after a detection.
In our model, we regard dead time as a shift between the detection event and the external time reference.
This shift does not affect the randomness of the raw data.

\item
The probability for multi-photon emission from an attenuated CW laser is nonzero. Since we set the time reference period to be smaller than the
detector dead time of 45 ns, we can get at most one detection in a period $(t,t+T)$.
When $k$ photons appear in a period, given that an ideal detector is used,
detection events at the small time bins $\{\tau_{n_1},\tau_{n_2},\dots ,\tau_{n_k}\}$ will be
announced. However, in the experiment only the first detection event $\tau_{\hat{n}}$ is recorded as the raw data,
where $\hat{n}=\min\{n_1,n_2,\dots ,n_k\}$. Therefore,
for a detection event the conditional probability of getting result $\hat{n}=i$ given that $k$ photons appear in a period is
\begin{equation} \label{P(n=i|k)}
\begin{aligned}
P(\hat{n}=i|k)=(1-\frac{i-1}{N_b})^k-(1-\frac{i}{N_b})^k,
\end{aligned}
\end{equation}
where $i=\{1,2,\dots,N_b\}$.

\end{enumerate}

The randomness of the raw data is evaluated by min-entropy, which is defined as
\begin{equation} \label{min-entropy}
\begin{aligned}
H_\infty &= -\log(\max P_i),
\end{aligned}
\end{equation}
where $P_i$ is the detection event probability in time bin $\tau_i$.
Thus, we need to figure out the detection event with the highest probability. From Eq.~\eqref{P(n=i|k)} one can easily see that $P(n=1|k)$
is the largest for any $k$. The upper bound of $P_1$ is given by
\begin{equation} \label{p1}
\begin{aligned}
P_1&=\frac{1}{1-e^{-\lambda T \eta}}\sum_{k=1}^\infty P(n=1|k)P(k)\\
&=\frac{1}{1-e^{-\lambda T \eta}}\sum_{k=1}^\infty\frac{(\lambda T\eta)^k e^{-\lambda T \eta}}{k!}[1-(1-\frac{1}{N_b})^k]\\
&\le\frac{1}{1-e^{-\lambda T \eta}}\sum_{k=1}^\infty\frac{(\lambda T\eta)^k e^{-\lambda T \eta}}{k!}\frac{k}{N_b}\\
&=\frac{\lambda T \eta}{N_b(1-e^{-\lambda T \eta})},\\
\end{aligned}
\end{equation}
where $1/(1-e^{-\lambda T \eta} )$ is the normalizing factor of $P(k)$. Then, the lower bound of min-entropy in this scheme is given by
\begin{equation}
\label{min-entropy dead time}
\begin{aligned}
H_\infty &= -\log(\max P_i)=-\log(P_1) \\
&\ge \log N_b+\log(1-e^{-\lambda T \eta})-\log(\lambda T \eta).
\end{aligned}
\end{equation}
One can see that the min-entropy depends on the experimental parameter of $\lambda T \eta$ that is related to the SPAD count rate.

\begin{figure}[tb]
\centering
\includegraphics[width=8 cm]{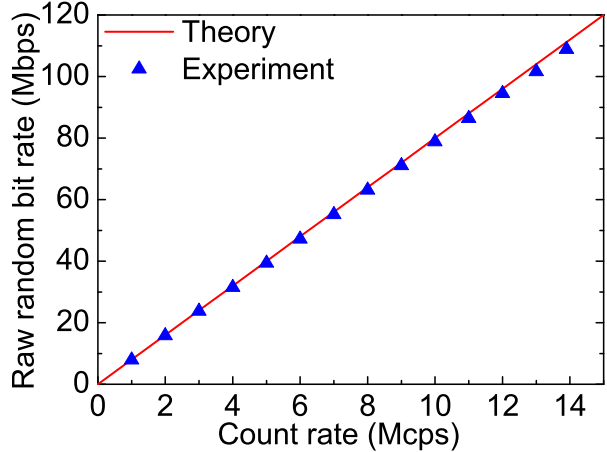}
\caption{Raw random bit rate as a function of the SPAD count rate.} \label{fig3}
\end{figure}

We measure the raw random bit rates at different SPAD count rates. As shown in Fig.~\ref{fig3}, the random bit rate linearly increases with SPAD count rate up to 109 Mbps at the detector's saturation rate of 13.9 Mcps. The minor deviations between experimental and theoretical values are mainly caused by data loss during the process of timing measurement, in which the TDC chip periodically requires a partial reset for continuous operation.

When the SPAD count rate is 13.9 Mcps, the parameter of $\lambda T \eta$ is 1.52. We substitute the parameters into the Eq.~\eqref{min-entropy dead time} and estimate the min-entropy of raw data to be $0.88$ per bit. Then, we apply a $universal_{2}$ hash function, Toeplitz matrix, to extract final random bits \cite{XQMXZL12,MXXQQL13}. The Toeplitz matrix with a size of $n\times m$ can extract an $m$-bit uniform string from $n$ bit raw data if the min-entropy is greater than $m/n$.
In our data post-processing, we take $n=3.36\times10^{7}$ and $m=2.95\times10^{7}$.
Software implementation of Toeplitz matrix hashing based on fast Fourier transform (FFT) is applied for randomness extraction. After the randomness extraction, the final random bit rate reaches 96 Mbps.

Our QRNG system can run stably at a raw bit rate of 109 Mbps, and over 1 Tb random bits are acquired after randomness extraction in the experiment.
We employ the NIST statistical test suite \cite{NIST10} to assess the randomness of the final data,
where 30 data files of 1 Gb size each are randomly selected for testing and all the files pass the NIST tests.
One of the test results is shown in Table \ref{tab:T1}. We also test the random data files with different sizes generated at the different SPAD count rates and all of them pass the NIST tests.

\begin{table}[t]
\tabcolsep0.02in
\caption{\label{tab:T1}Typical test result of 1 Gb random bits. In the tests that produce multiple outcomes of p-values and proportions, the worst outcomes are selected. Since the significance level in the tests is $\alpha = 0.01$, the p-value should be larger than 0.01 and the proportion should be above 0.98. }
\begin{tabular}{l|lcl}
\hline
\hline
Statistical test &P-value &Proportion &Result\\
\hline
Frequency                  &0.329850   &0.994  &Pass\\
Block Frequency            &0.194813   &0.988  &Pass\\
Cumulative Sum             &0.490483   &0.994  &Pass\\
Runs                       &0.366918   &0.990  &Pass\\
Longest Run                &0.368587   &0.987  &Pass\\
Rank                       &0.701366   &0.986  &Pass\\
FFT                        &0.735908   &0.988  &Pass\\
Non Overlapping Template   &0.017912   &0.988  &Pass\\
Overlapping Template       &0.352107   &0.985  &Pass\\
Universal                  &0.834308   &0.988  &Pass\\
Approximate Entropy        &0.881662   &0.989  &Pass\\
Random Excursions          &0.106666   &0.987  &Pass\\
Random Excursions Variant  &0.173452   &0.992  &Pass\\
Serial                     &0.189625   &0.988  &Pass\\
Linear Complexity          &0.222480   &0.992  &Pass\\
\hline
\hline
\end{tabular}
\end{table}

To improve the random bit generation rate of our QRNG system, one can use higher count rate detectors and higher time resolution TDCs.
For example, a silicon photon multiplier with maximum count rate of 430 Mcps \cite{ELRZG07} along with a high-precision TDC with 1 ps time resolution \cite{TDC08} may yield a generation rate of 4 Gbps.

In summary, we design and test a practical high-speed QRNG based on the photon arrival time from a CW laser. High min-entropy raw data are generated from the timing measurement of single-photon detection relative to an external time reference. Compared with the previous photon arrival time based QRNG schemes, our scheme can significantly eliminate the bias existing in the raw data and generate almost uniformly distributed raw bits. We also model the generator and evaluate the min-entropy of the raw data, by taking into account imperfections of the QRNG, such as the detection efficiency, dark count, dead time of the single-photon detector, and the multi-photon emission from the laser source. In the experiment, the maximum raw random bit rate reaches 109 Mbps. After randomness extraction with FFT-based Toeplitz matrix hashing, the final random bits reach a rate of 96 Mbps and are able to pass the standard randomness tests.
The simplicity and robustness of the QRNG scheme allows its applications in various practical situations.

We acknowledge insightful discussions with W.~Wang. This work has been financially supported by the National Basic Research Program of China Grant No.~2013CB336800, the National High-Tech R\&D Program Grant No.~2011AA010802, the National Natural Science Foundation of China Grant No.~61275121, and the Chinese Academy of Sciences. X. M. and Z. Z. acknowledge the financial support from the National Basic Research Program of China Grants No.~2011CBA00300 and No. 2011CBA00301.

\end{document}